\documentclass[12pt]{iopart}

\usepackage{iopams}  
\usepackage{color}
\usepackage{dsfont}
\usepackage{cite}
\usepackage{graphicx}
\begin{document}

\def\comment#1{{\color{red}\textbf{#1}\color{black}}}
\def\quabla{\square}
\def\del{\partial}
\def\text#1{\mbox{#1}}
\def\xvec{\mathbf{x}}
\def\yvec{\mathbf{y}}
\def\zvec{\mathbf{z}}
\def\arccosh{\text{arccosh}}
\def\d{\text{d}}
\def\id{\mathds{1}}
\parskip 2mm

\title
[Fractional Klein-Gordon Equation on AdS$_{2+1}$]
{Fractional Klein-Gordon Equation on AdS$_{2+1}$}

\author{Pablo Basteiro, Janine Elfert, Johanna Erdmenger, and\\ Haye Hinrichsen}

\address{Faculty for Physics and Astronomy, University of W\"urzburg\\
Am Hubland, 97074 W\"urzburg, Germany}
\ead{hinrichsen@physik.uni-wuerzburg.de}
\vspace{10pt}

\begin{abstract}
We propose a covariant definition of the fractional Klein-Gordon
equation with long-range interactions independent of the metric of the
underlying manifold. As an example we consider the fractional
Klein-Gordon equation on AdS$_{2+1}$, computing the explicit kernel
representation of the fractional Laplace-Beltrami operator as well as
the two-point propagator of the fractional Klein-Gordon equation. Our
results suggest that the propagator only exists if the mass is 
small compared to the inverse AdS radius, presumably because the AdS space expands 
faster with distance as a flat space of the same dimension.
Our results are expected to be useful in particular for new applications of the
AdS/CFT correspondence within statistical mechanics and quantum information.\footnote{Submitted to J. Phys. A.:\\Fine Latticework: Celebrating the Craftsmanship of Robert M. Ziff in Honour of his 70th Birthday}
\end{abstract} 

\section{Introduction}

In quantum field theory and statistical physics, elementary interactions are usually assumed to be local. For example, Lagrange densities are typically defined in terms of local expressions in the fields and their derivatives, where the derivatives can be understood as operators which couple infinitesimally neighbored field values. Similarly, in statistical physics the most lattice modes are defined in terms of nearest-neighbor interactions.

The premise of local interactions does not necessarily imply that the resulting physics is local as well. For example, the electromagnetic field, defined by a local Lagrangian, exhibits Coulomb forces whose range is infinite. Further examples include critical phenomena in statistical physics and conformal field theories which exhibit long-range correlations with power-law characteristics.

Although in most physical situations the assumption of local interactions appears to be natural, it is interesting to study what changes if we allow the elementary interactions to become non-local. This question has been investigated for long time and has led to a large variety of studies in different contexts (for a review see e.g.~\cite{campa2014physics}). The impact of non-local interactions has been studied particularly thoroughly in the context of phase transitions, critical phenomena and conformal field theories. Although the results are so diverse that they cannot be summarized in a few lines, a generic picture emerges which can be sketched as follows:
\begin{itemize}
\item 
    Systems with non-local interactions extending over a \textit{finite} range or with \textit{exponentially} decaying coupling strength behave on large scales as if they were local. This includes, for example, next-nearest neighbor interactions in lattice models or the addition of irrelevant higher-order derivatives in quantum field theories. For instance, the phase transition in an Ising model with non-local but exponentially decaying coupling constants still belongs to the Ising universality class.
\item 
    Non-local interactions with long-range couplings decaying asymptotically as a power law $s^{-\kappa}$ with the distance $s$ are capable to entirely change the macroscopic properties of a physical system. For example, an Ising model with non-local algebraically decaying coupling constants may exhibit a different type of universal behavior on large scales.
\end{itemize}
In many of these studies algebraically decaying couplings are chosen because this choice allows one to introduce genuine long-range interactions without imposing a particular scale. Therefore, when studying scale-invariant phenomena such as second-order phase transitions or conformal field theories, one expects that scale invariance is not broken by introducing such long-range interactions.

The impact of long-range couplings decaying as $s^{-\kappa}$ depends of course on the value of the exponent $\kappa$, which plays here the role of an additional control parameter. Therefore, it is interesting to study how the macroscopic properties of a system change if this exponent is varied. Clearly, for $\kappa=0$ the system is globally coupled, resulting in some kind of mean-field behavior. On the other hand, taking $\kappa \to \infty$, the interaction strength decays so rapidly that we expect to recover the short-range behavior with local interactions. Between these two extremes such models are expected to exhibit an interesting crossover behavior.
This crossover regime has been analyzed in a large variety of systems in different contexts, including anomalous diffusion~\cite{metzler2000random}, L\'evy flights~\cite{fogedby1994langevin,chaves1998fractional}, equilibrium critical phenomena~\cite{luijten2002boundary,blanchard2013influence}, non-equilibrium phase transitions and driven systems~\cite{hinrichsen2007non,bouchet2010thermodynamics}, epidemic spreading~\cite{janssen1999levy,linder2008long}, as well as field-theoretic models~\cite{brezin2014crossover}. Most of these studies confirm that the crossover between the short-range and the mean-field regime is observed only in a finite range of the control exponent, where one observes a $\kappa$-dependent universal behavior with continuously varying critical exponents.

Algebraically decaying long-range interactions are closely related to a special topic in mathematics called \textit{fractional calculus} \cite{oldham1974fractional,gorenflo1997fractional,hilfer2000applications,baleanu2012fractional} which deals with non-integer derivatives and integrals~\cite{hilfer2008threefold}. This relation is due to the fact that fractional derivatives can be represented by linear integral kernels with power-law characteristics. Therefore, as a rough guideline, if a short-range model is defined in a continuum description in terms of ordinary derivatives, then the corresponding long-range variant can be obtained by simply replacing ordinary derivatives with fractional ones.

The vast majority of studies on algebraically decaying long-range interactions are concerned with non-relativistic systems in flat geometries~(for a review see e.g.~\cite{gupta2017world}). Considerable less studies are available for   relativistic systems such as the fractional wave equation \cite{schneider1989fractional} and its massive counterpart, the fractional Klein-Gordon equation~\cite{vong2014compact,karaagac2019study,singh2019reliable}. The same applies to studies of systems with fractional long-range interactions in curved geometries. Therefore, it is interesting to investigate the question whether it is possible to implement long-range interactions by fractional derivatives in relativistic systems on a curved background in a covariant way and to analyze their properties.

In this paper we consider the example of a fractional Klein-Gordon
equation living on curved space. As a curved background, it is natural
to choose a spacetime with maximal symmetry and constant
curvature. However, a compact space with positive curvature such as a
sphere with a finite circumference would truncate the interaction
range. For this reason it is an obvious choice to study a maximally symmetric space with negative curvature, an Anti-de-Sitter (AdS) space. 

For reader with a statistical mechanics background, we note that AdS space plays an important role in many areas of physics and
mathematics. In particular, it is central to the AdS/CFT
correspondence~\cite{Maldacena:1997re,Witten:1998qj,Gubser:1998bc,magoo}
(also known as holography), which originates from string theory and
provides  a new perspective on quantum gravity. The correspondence
relates weakly coupled gravitational theories in $(d+1)$-dimensional
AdS space with strongly coupled conformal field theories (CFTs) defined on
the $d$-dimensional boundary of AdS. At the heart of the correspondence
 is a particular matrix large $N$ limit, in which
$N$ is taken to be large for the CFT which is typically a $SU(N)$
non-abelian gauge theory. 
The AdS/CFT correspondence has
allowed for a plethora of progress to be made with regard to
developing a consistent quantum theory of gravity. Moreover, it has
provided new insight into specific areas of physics, such as
phase transitions \cite{Witten:1998zw}, strongly coupled condensed
matter systems \cite{Hartnoll:2009sz}, quantum information theory
\cite{RevModPhys.88.015002} and cosmology \cite{Fischler:1998st} (see
\cite{ammon2015gauge} for more applications).

It is expected that in the large $N$ limit given above, the AdS/CFT
correspondence satisfies the property of {\it bulk locality} according
to which the bulk gravity theory is local \cite{Heemskerk,Terashima2021,CaronHuot2021},
also in other string theory parameter regions as for instance for
higher spin theories \cite{Bekaert:2015tva}. However, non-localities
may potentially arise when venturing beyond the large $N$ limit, a
parameter region hard to explore so far since it requires to formulate
the AdS/CFT correspondence for quantized string theories. Moreover,
the framework of AdS/CFT is constantly being expanded with new ideas
of interdisciplinary nature, and new gravity duals to statistical
mechanics systems in which non-locality plays are role are conceivable.
Here, we take a new step in this direction by considering fractional
equations of motion for a scalar field in AdS space. We believe our
results can contribute both to considering 1/N corrections within
AdS/CFT as well as to new applications of AdS/CFT to statistical mechanics.
In view of these potential applications, we refer to the recent work \cite{PhysRevA.102.032208,lenggenhager2021electriccircuit}
on discrete hyperbolic spaces in the context of condensed matter physics.

Instances of fractional calculus within the AdS/CFT correspondence are
scarce \cite{PhysRevD.94.126018,LaNave:2017nex}, but they have the common topic of introducing non-locality in the boundary theory. In the work \cite{PhysRevD.94.126018,LaNave:2017nex}, bulk locality is still preserved, but bulk $p$-forms of a particular mass are shown to be dual to a non-local operator in the CFT, namely the fractional Laplacian on the boundary CFT on a flat background. We instead introduce the fractional Laplacian as a non-local operator in the bulk, i.e. on a curved background. 

In order to implement long-range interactions on curved spaces in a physically meaningful way, we propose a method to ``fractionalize'' a differential equation of motion covariantly independent of the background metric. We demonstrate this approach for the scalar Klein-Gordon equation and its massless special case, the wave equation, living in AdS space. For simplicity, we decided to restrict the analysis to ${2+1}$ dimensions throughout the whole paper although similar calculations can be done in any dimension.

One should keep in mind that the AdS space, although being non-compact, singles out a particular scale, namely, the AdS radius $L$. Therefore, all results are expected to reflect this scale in the formulas. As a crosscheck, taking $L$ to infinity, we should be able to recover previously known results in flat Minkowski spacetime.

Our main results are the following:
\begin{itemize}
\item[-] 
We propose a simple method to introduce long-range interactions covariantly independent of the background metric.
\item[-]
For the special case of the scalar Klein-Gordon equation in AdS$_{2+1}$ we derive an explicit kernel representation of the fractional operator which allows us to read off the actual interaction strength as a function of the geodesic distance. 
\item[-]
We find an explicit expression for the two-point propagator (Green's function) of the fractional wave equation on AdS$_{2+1}$ as well as a series expansion for the massive propagator of the fractional Klein-Gordon equation. 
\item[-]
Finally, we verify numerically that the differential equation in its kernel repre\-sentation applied to the propagator exhibits the desired properties.
\end{itemize}
%

\section{Ordinary Klein-Gordon equation on AdS$_{2+1}$}
%
Setting up a fractional equation of motion for a relativistic system, we have to choose a covariant approach, that is, the equation should be invariant under diffeomorphisms which generalize Lorentz transformations in curved spaces. The easiest way to achieve this goal is to reformulate the problem solely in terms of geodesic distances. Since geodesic distances are covariant by themselves, this ensures the covariance of the whole approach. In this section we demonstrate how this can be done in the case of the scalar Klein-Gordon equation (KGE) on AdS$_{2+1}$.

\subsection{Ordinary Klein-Gordon equation in flat Minkowski space}
Before starting let us fix some notations by briefly recalling the situation  in a flat (2+1)-dimensional Minkowski space. Here the Klein-Gordon equation is given by
\begin{equation}
(\quabla -m^2 )\phi(\xvec) \;=\; 0\,,
\end{equation}
where $\xvec=(t,x,y)$ denotes the position vector, $\phi(\xvec)$ is a real-valued scalar field, $m\geq0$ is the mass and $\quabla=\eta^{\mu\nu}\del_\mu \del_\nu=-\del_t^2+\del_x^2+\del_y^2$ is the d'Alembert operator. Throughout this paper we use the \textit{mostly-plus} convention $\eta_{\mu\nu}=\text{diag}(-1,1,1)$ which leads to the minus sign in front of the squared mass.

In what follows we are concerned with solutions of the KGE which depend solely on the relativistic distance $s=\sqrt{x_\mu x^\mu}=\sqrt{-t^2+x^2+y^2}$ from the origin, analogous to radially symmetric functions in Euclidean spaces. The d'Alembert operator acts on such functions as $\quabla f(s) = f''(s) + \frac{2}{s}f'(s)$, turning the KGE into
\begin{equation}
\label{NonfractionalKGflat}
\phi''(s) + \frac 2 s \phi'(s) -m^2 \phi(s) \;=\; 0
\end{equation}
which can also be written in an equivalent factorized form as
\begin{equation}
\label{NonfractionalKGflatFac}
\frac{1}{s^2} \partial_s\, s^2 \partial_s\, \phi(s) -m^2 \phi(s) \;=\; 0\,.
\end{equation}
For $s \neq 0$ the two fundamental solutions of the homogeneous KGE are
\begin{equation}
\label{HomogeneousSolutions}
\phi_\pm(s) = \frac{e^{\pm m s}}{4 \pi s}\,.
\end{equation}
In practice, it is often simpler to consider the analytic continuation of the Klein Gordon equation to imaginary time $\tau:=i t$, where $s=\sqrt{\tau^2+x^2+y^2}$ is the Euclidean distance and $\quabla=\del_\tau^2+\del_x^2+\del_y^2$ is just the ordinary 3D Laplacian. Note that in this case the equations (\ref{NonfractionalKGflat})-(\ref{HomogeneousSolutions}) remain valid, that is, the radial representation is correct in the Lorentzian as well as in the Euclidean setting.

For solving the inhomogeneous KGE, we are especially interested in the Green's function defined by
\begin{equation}
(\quabla -m^2 )G(\xvec) \;=\; \delta^3(\xvec)\,.
\end{equation}
which translates into
\begin{equation}
G''(s) + \frac 2 s G'(s) -m^2 G(s) \;=\; \frac{\delta(s)}{2\pi s^2}\,.
\end{equation}
In the Euclidean case the normalizable solution reads
\begin{equation}
\label{nonfracgreensEucli}
G(s)\;=\;-\frac{e^{-ms}}{4 \pi s}\qquad\text{for } s>0 \, .
\end{equation}
In Minkowski space one obtains the same Green's function with an imaginary factor in front. 

\subsection{Ordinary Klein-Gordon equation on AdS$_{2+1}$ in terms of geodesic distances}
%
Let us now turn to the (non-fractional) KGE on an Anti-de-Sitter (AdS) space in 2+1 dimensions. The AdS$_{2+1}$ space is defined as a maximally symmetric Riemann manifold with negative curvature and can be represented in various coordinate systems~\cite{bayona2007anti}. In what follows we use Poincar{\'e} patch coordinates $(t,x,z)$ with the metric
\begin{equation}
g_{\mu\nu}=\frac{L^2}{z^2}\,\text{diag}(-1,1,1)\,,\quad
g^{\mu\nu}=\frac{z^2}{L^2}\,\text{diag}(-1,1,1)\,,\quad
\sqrt{|g|}=\frac{L^3}{z^3}\,,
\end{equation}
where $L>0$ is the AdS radius. In these coordinates the Laplace-Beltrami operator, which generalizes the d'Alembert operator in a curved space, reads
\begin{equation}
\label{laplacebeltrami}
\quabla_g \;=\; \frac{1}{\sqrt{|g|}}\del_\mu \sqrt{|g|} g^{\mu\nu} \del_\nu  \;=\;
\frac{z^2}{L^2}\Bigl( -\del_t^2+\del_x^2+\del_z^2-\frac{1}{z}\del_z \Bigr)\,.
\end{equation}
The \textit{chordal distance} $\xi(\xvec,\xvec')$ between two points $\xvec=(t,x,z)$ and $\xvec'=(t',x',z')$ (i.e. the straight-line distance in the embedding space of the AdS hyperboloid) is defined as
\begin{equation}
\xi(\xvec,\xvec') \;=\; \frac{2 z z'}{z^2+{z'}^2-(t-t')^2+(x-x')^2}\,.
\end{equation}
The \textit{geodesic distance} $s(\xvec,\xvec')$ between two points $\xvec$ and $\xvec'$, defined as the length of the geodesic line within the AdS manifold, can be expressed in terms of the chordal distance via
\begin{equation}
s(\xvec,\xvec') \;=\; L\, \arccosh\Bigl(\frac 1{\xi(\xvec,\xvec')}\Bigr)\,.
\end{equation}
Since the AdS space is homogeneous (i.e. it looks the same from all points), we can  declare the point
\begin{equation}
x'=t'=0\,,\quad z'=1
\end{equation}
as the origin of the coordinate system. Again we are interested in ``radially symmetric'' functions $f$ which depend only on the geodesic distance from the origin:
\begin{equation}
f(t,x,z) \;\equiv\; f(s) \;=\; f\Bigl( L \, \arccosh\Bigl( \frac{1+z^2-t^2+x^2)}{2z} \Bigr) \Bigr)\,.
\end{equation}
If the Laplace-Beltrami operator acts on such a function, its action can be rewritten as
\begin{equation}
\label{radialLaplaceBeltrami}
\quabla_g f(s) \;=\; \Bigl[\del_s^2+\frac{2}{L}\coth\Bigl( \frac{s}{L} \Bigr)\del_s \Bigr] f(s)\,,
\end{equation}
turning the KGE into
\begin{equation}
\label{NonfractionalKGonAdS}
\phi''(s) + \frac{2}{L}\coth\Bigl( \frac{s}{L} \Bigr) \phi'(s) -m^2 \phi(s) \;=\; 0\,
\end{equation}
or in the factorized form
\begin{equation}
\label{NonfractionalKGonAdSFac}
\frac{1}{ {\sinh^2\left(\frac{s}{L}\right)}}\,\partial_s \,\sinh^2\left(\frac{s}{L}\right)\, \partial_s \, \phi(s) -m^2 \phi(s) \;=\; 0\, .
\end{equation}
As can be seen, taking $L\to\infty$ this equation reduces to the KGE~(\ref{NonfractionalKGflat}) in a flat space. The same equation is obtained in the Euclidean case if $t$ is replaced by $\tau=it$.
%
%
\subsection{Green's function of the ordinary Klein-Gordon equation on AdS$_{2+1}$}
%
To find the Green's function of the short-range KGE on AdS$_{2+1}$, we first express the defining inhomogeneous differential equation 
\begin{equation}
(\quabla_g-m^2)G(\xvec)\;=\;\frac{\delta^3(\xvec)}{\sqrt{|g|}}
\end{equation}
in terms of the geodesic distance~$s$. Since the r.h.s. contributes only at the origin where $\sqrt{|g|}=1$, we can equivalently study the equation $(\quabla_g-m^2)G(\xvec)=\delta^3(\xvec)$. Moreover, since volume integrals in the Euclidean AdS$_{2+1}$ can be expressed as 
\begin{equation}
\int_{-\infty}^{+\infty}\d^3x \,f(\xvec)=4\pi L^2 \int_0^\infty \d s \, \sinh^2\bigl( \frac s L \bigr) f(s)
\end{equation}
it is clear that $\delta^3(\xvec)=\frac{\delta(s)}{2\pi L^2 \sinh^2(s/L)}$. Thus in terms of geodesic distances the Green's function $G(s)$ is defined by the the inhomogeneous equation
\begin{equation}
G''(s)+\frac{2}{L}\coth\Bigl(\frac{s}{L}\Bigr) G'(s)-m^2 G(s) \;=\; \frac{\delta(s)}{2\pi L^2 \sinh^2\bigl(\frac{s}{L}\bigr)}\,.
\end{equation}
This equation has two solutions. Only one of them is normalizable, being given by
\begin{equation}
\label{nonfractionalMassiveGreens}
G(s) \;=\; -\frac{\exp\Bigl( -\frac{s}{L} \sqrt{1+L^2m^2} \Bigr)}{4\pi L \sinh\bigl( \frac s L \bigr)}\,.
\end{equation}
In the limit $L\to \infty$ this solution consistently reduces to~(\ref{nonfracgreensEucli}). Note that this solution is compatible with existing results in the literature (see~\ref{AppendixEquivalence}) and that it is formally valid even for negative squared mass above the so-called Breitenloher-Freedman bound $m^2 > -L^{-2}$~\cite{breitenlohner1982positive}. 

\section{Fractional Klein-Gordon equation on AdS$_{2+1}$}

In this section we turn to the construction of a fractional version of the KGE. As a conceptual problem it turns out that the resulting equation depends crucially on the specific way in which the long-range interactions are implemented. Naively the simplest approach would be to just replace ordinary with fractional derivatives. However, we observed that in a curved space the resulting equation depends on the specific representation in which this replacement is made. Given this non-uniqueness, which of these variants is physically reasonable? In order to avoid this ambiguity, we will adopt a scheme suggested by Bochner~\cite{bochner1949diffusion} which allows us to define a fractional Laplacian independent of the structure of the underlying manifold. 

\subsection{Fractional derivatives}

Before starting, let us briefly summarize some basic facts about fractional derivatives. In the literature one finds a confusing variety of definitions~\cite{oldham1974fractional} which differ in their symmetry properties and function spaces. In practical terms, fractional derivatives are most easily introduced by looking at their action in Fourier space. There are mainly two classes of fractional derivatives. The first one comprises \textit{directed} fractional derivatives which act in Fourier space as
\begin{equation}
\frac{\d^\alpha}{\d t^\alpha} \, e^{i \omega t} = (i \omega)^{\alpha}\, e^{i \omega t} \,,
\end{equation}
where $\alpha>0$ is a control exponent. These variants generalize the first derivative (the generator of translations) and are often used to implement algebraically distributed waiting times in statistical mechanics~\cite{fogedby1994langevin,adamek2005epidemic}. The second class, on which we will focus in the present work, comprises \textit{isotropic} fractional derivatives which act on a plane wave according to
\begin{equation}
\Delta^{\alpha} \, e^{i \vec k \cdot \vec x} = -(k^2)^{\alpha}\, e^{i \vec k \cdot \vec x} \,.
\end{equation}
These derivatives generalize the second-order Laplacian (the generator of diffusion) and are often used to model isotropic L\'evy flights in space~\cite{fogedby1994levy}. Similarly, in the KGE, we can define a fractional version of the d'Alembert operator in a relativistic setting via
\begin{equation}
\label{fourier}
\quabla^\alpha \, e^{i k_\mu x^\mu} \;=\; -(k_\mu k^\mu)^\alpha \, e^{i k_\mu x^\mu}\,,
\end{equation}
where $\alpha>0$. By definition this operator is linear and obeys the composition law
\begin{equation}
\label{CompositionLaw}
\quabla^\alpha \quabla^\beta\;=\;-\quabla^{\alpha+\beta}\,.
\end{equation}
Moreover, we expect to retrieve the usual short-range d'Alembertian in the limit $\alpha\to 1$. 

As any linear operator, the fractional d'Alembert operator can be represented by an integral kernel. As outlined in~\cite{kwasnicki2017ten}, there are several equivalent integral representations. One of them is the so-called singular integral representation in $d$ dimensions:
\begin{equation}
\label{singular}
\quabla^\alpha\,f(\xvec) \;=\; -\frac{4^\alpha \Gamma\bigl(\frac{d+1}{2}+\alpha\bigr)}{\pi^{\frac{d+1}{2}} \Gamma(-\alpha)}\int \d^{d+1} z \,\frac{f(\xvec+\zvec)-f(\xvec)}{\bigl(z_\mu z^\mu\bigr)^{\frac{d+1}{2}+\alpha}}\,.
\end{equation}
This representation is important because it gives us a direct intuition, namely, it can be interpreted as a non-local transport process (L\'evy flight) from the position $\xvec$ to $\xvec+\zvec$ weighted by the relativistic distance $s=\sqrt{z_\mu z^\mu}$ raised to the power $d+1+2\alpha$. Thus, by replacing an ordinary by a fractional d'Alembert operator we can effectively implement long-range interactions with power-law characteristics.

\subsection{Bochner's approach to fractional derivatives}

On curved manifolds, it is not so obvious how to define a fractional version of the Laplace-Beltrami operator defined in~(\ref{laplacebeltrami}) since Fourier techniques are no longer available. 

One possibility would be to simply replace ordinary by fractional derivatives. We first tried this in Minkowski space using the factorized form of the KGE in Eq.~(\ref{NonfractionalKGflatFac}). More specifically, we replaced the partial derivatives by fractional Gerasimov-Caputo derivatives because they return zero on constant functions and reproduce ordinary derivatives in a certain limit. We calculated the corresponding kernel representation and applied it to a exponential functions and powers, getting consistent results. However, we could not reproduce the concatenation law~(\ref{CompositionLaw}). On AdS$_{2+1}$, the implementation and the physical meaning of this approach was even less clear.

Therefore, instead of guessing replacement rules, one needs a physically consistent and transparent way for defining a fractional Laplace-Beltrami operator which does not depend on the structure of the underlying manifold. In this paper we suggest to adopt an approach that goes back to S. Bochner~\cite{bochner1949diffusion,kwasnicki2017ten}. Accordingly, any linear operator $\mathbf{L}$ acting on some function space can be associated with a fractional version $\mathbf{L}^\alpha $ defined by
\begin{equation}
\label{BochnerDefinition}
\mathbf{L}^\alpha \;:=\;-\frac{1}{\Gamma(-\alpha)}\int_0^\infty \d \tau  \,\tau^{-1-\alpha}\, \bigl( e^{\tau \mathbf{L}} - \id\bigl)\,.\qquad\qquad0<\alpha<1
\end{equation}
In the literature the operator $\mathbf{K}_\tau=e^{\tau\mathbf{L}}$ is known as the \textit{heat kernel} of $\mathbf{L}$ because $\mathbf{K}_\tau$ is the formal solution of the generalized heat conduction equation $\frac{\d}{\d\tau}\mathbf{K}_\tau = \mathbf{L}\mathbf{K}_\tau$.

As can be easily proven, Bochner's definition nicely reproduces the short-range operator in the limit $\alpha\to 1$ on a suitable function space. In this limit the prefactor in front of the integral goes to zero as $1-\alpha$, compensated by an increasingly diverging contribution of the integral for small $\tau$, governed by the first-order contribution in the Taylor expansion of  $( e^{\tau \mathbf{L}} - \id)$.

Note that Bochner's representation is completely independent of the dimension and the metric of the manifold. Moreover, its interpretation is very transparent from the physical point of view. Suppose that $\mathbf L$ is the generator of some physical process, let it be translation or diffusion. Then this process is carried out by exponentiation over some ``distance'' $\tau$, and the result is weighted by a fractional power of this distance. 

\subsection{Fractional Klein-Gordon equation in AdS$_{2+1}$}
%
If we apply Bochner's definition~(\ref{BochnerDefinition}) to the d'Alembert operator in a flat Minkowski space we get the fractional operator
\begin{equation}
\label{bochner}
\quabla^\alpha\,f(\xvec) \;=\; 
-\frac{1}{\Gamma(-\alpha)}\,
\int_0^\infty \d \tau\,\tau^{-1-\alpha}\, \bigl( e^{\tau\quabla}-1) f(\xvec)\,.
\end{equation}
Inserting a plane wave $f(\xvec)=e^{i k_\mu x^\mu}$ with $\square f(\xvec)=-k_\mu k^\mu f(\xvec)$ it is easy to check that
\begin{equation}
\quabla^\alpha\,e^{i k_\mu x^\mu} =
-\frac{e^{i k_\mu x^\mu}}{\Gamma(-\alpha)}\,
\underbrace{\int_0^\infty \d \tau\,\tau^{-1-\alpha}\, \bigl( e^{-\tau k_\mu k^\mu}-1)}_{=\Gamma(-\alpha)(k_\mu k^\mu)^\alpha} =
- (k_\mu k^\mu)^\alpha\,e^{i k_\mu x^\mu},
\end{equation}
hence the definition is compatible with~(\ref{fourier}) and~(\ref{singular}). This encourages us to \textit{define} the fractional Laplace-Beltrami operator $\quabla_g^\alpha$ on AdS$_{2+1}$ in the same way as
\begin{equation}
\label{fractionalLaplaceBeltrami}
\quabla_g^\alpha \;:=\;-\frac{1}{\Gamma(-\alpha)}\int_0^\infty \d \tau  \,\tau^{-1-\alpha}\, \bigl( e^{\tau \quabla_g} - \id\bigl)\,,
\end{equation}
where $\quabla_g$ is the non-fractional Laplace-Beltrami operator defined in~(\ref{laplacebeltrami}) and~(\ref{radialLaplaceBeltrami}) and where $0<\alpha<1$ is the factional control parameter. More specifically, if we apply this operator to a function $f$, the above definition reads
\begin{equation}
\label{explicitFractionalLaplaceBeltrami}
[\quabla_g^\alpha f](\xvec) \;=\;-\frac{1}{\Gamma(-\alpha)}\int_0^\infty \d \tau  \,\tau^{-1-\alpha}\, \Bigl( \bigl[e^{\tau \quabla_g}f\bigr](\xvec) - f(\xvec)\Bigr)\,.
\end{equation}
As argued above, we expect to retrieve the short-range operator the limit $\alpha\to 1$, that is $\quabla_g = \lim_{\alpha\to 1}\quabla_g^\alpha$. Moreover, one can show that this operator is compatible with the composition law $\quabla^\alpha \quabla^\beta\;=\;-\quabla^{\alpha+\beta}$, allowing one to extend the range to $\alpha>1$.

\section{Kernel representation}
\subsection{Kernel representation in geodesic distances}
%
As any linear operator, the heat kernel $\mathbf{K}_\tau=e^{\tau \quabla_g}$ in the expression above can be represented as an integral
\begin{equation}
\label{kernelRepresentation}
 [e^{\tau \quabla_g} f] (\xvec) \;=\; \int \d^3 y \,\sqrt{|g|}\, K_\tau(\xvec,\yvec)\, f(\yvec)
\end{equation}
with a suitable kernel function $K_\tau(\xvec,\yvec)$ to be determined. Assuming that this kernel function depends only on the geodesic distance $s(\xvec,\yvec)$ between the points $\xvec$ and~$\yvec$ and likewise that $f$ depends only on the geodesic distance from the origin $s(\yvec)$, i.e.,
\begin{equation}
\label{radiallysymmetric}
K_\tau(\xvec,\yvec) \equiv K_\tau\bigl(s(\xvec,\yvec)\bigr)\,,\qquad\quad
f \bigl(\yvec\bigr)\equiv f \bigl(s(\yvec)\bigr)\,,
\end{equation}
we expect the result $e^{\tau \quabla_g} f$ to exhibit the same type of symmetry. This allows us to reformulate the integral (\ref{kernelRepresentation}) solely in terms of geodesic distances $s=s(\xvec)$, $s'=s(\yvec)$, and $s''=s(\xvec,\yvec)$ (for details see \ref{AppendixKaellenFormula}):
\begin{equation}
\label{only21}
\fl 
\label{radialFractionalLaplaceBeltrami}
[e^{\tau \quabla_g} f ](s)\;=\;\frac{2\pi L}{\sinh\bigl( \frac{s}{L} \bigr)}\,\,\int_0^\infty \d s' \,\sinh\Bigl( \frac{s'}{L} \Bigr)\, f (s') \, \underbrace{\int_{|s-s'|}^{s+s'} \d s'' \,\sinh\Bigl( \frac{s''}{L} \Bigr)\,\, K_\tau(s'')}_{=: \,J_\tau(s,s')}\,.
\end{equation}
Thus the determination of the kernel representation essentially amounts to computing the inner integral, denoted as $J_\tau(s,s')$. Please keep in mind that the expression (\ref{only21}) is valid only in 2+1 dimensions.

Inserting the heat kernel~(\ref{radialFractionalLaplaceBeltrami}) into Eq.~(\ref{explicitFractionalLaplaceBeltrami}) and using that $\id = e^{\tau \quabla_g} \id$ we obtain
\begin{equation}
\fl
\eqalign{
[\quabla_g^\alpha f](s) &= -\frac{1}{\Gamma(-\alpha)}\int_0^\infty \d \tau  \,\tau^{-1-\alpha} \Bigl[ \Bigl( \frac{2\pi L}{\sinh\bigl( \frac{s}{L} \bigr)}\int_0^\infty \d s' \,\sinh\Bigl( \frac{s'}{L} \Bigr)\, J_\tau(s,s') f(s') \Bigr) - f(s)\Bigr] \\
&= -\frac{1}{\Gamma(-\alpha)}\int_0^\infty \d \tau  \,\tau^{-1-\alpha}\,\frac{2\pi L}{\sinh\bigl( \frac{s}{L} \bigr)}\int_0^\infty \d s' \,\sinh\Bigl( \frac{s'}{L} \Bigr)\, J_\tau(s,s') \Bigl(f(s')-f(s) \Bigr)\\
&= -\frac{2 \pi L}{\Gamma(-\alpha)}\int_0^\infty \d s' \frac{\sinh\bigl( \frac{s'}{L} \bigr)}{\sinh\bigl( \frac{s}{L} \bigr)}\underbrace{\int_0^\infty \d \tau  \,\tau^{-1-\alpha}J_\tau(s,s')}_{=:\,J(s,s')} \Bigl(f(s')-f(s) \Bigr)\,.
}
\end{equation}
%

\subsection{AdS heat kernel}
%
In order to evaluate the inner integral $J_\tau(s,s')$ in Eq.~(\ref{radialFractionalLaplaceBeltrami}) we need to know the kernel function $K_\tau(s)$ in AdS$_{2+1}$ as a function of the geodesic distance $s$ from the origin. Fortunately, such kernels have already been computed in Ref.~\cite{grigor1998heat}. Converting these existing results into the present notation, the heat kernel in AdS$_{2+1}$ is given by
\begin{equation}
\label{heatkernel21}
K_\tau(s) \;=\; \frac{1}{(4\pi \tau)^{3/2}}\,\frac{\frac{s}{L}}{\sinh\bigl( \frac s L\bigr) }\,  \exp\Bigl( -\frac{\tau}{L^2}- \frac{s^2}{4\tau} \Bigr)\,.
\end{equation}
In physical terms, this kernel describes how a point-like injection of heat diffuses in AdS$_{2+1}$ as a function of the geodesic distance $s$ over a fictitious time span $\tau$. As expected, in the flat-space limit $L\to \infty$ this expression reduces to a Gaussian. 

To convince ourselves that this expression is indeed correct, we check that it solves the heat conduction equation
\begin{equation}
\frac{\del}{\del\tau} \, K_\tau(s) \;=\; \quabla_g K_\tau(s) \;=\; \Bigl[\frac{2}{L}\coth\Bigl( \frac{s}{L} \Bigr)\del_s + \del_s^2\Bigr] K_\tau(s)
\end{equation}
and that its volume integral is correctly normalized for all $\tau>0$:
\begin{equation}
4\pi L^2 \int_0^\infty \d s \, \sinh^2\Bigl(\frac s L\Bigr) \, K_\tau(s) \;=\; 1\,.
\end{equation}
%
\subsection{Kernel representation of the fractional Laplace-Beltrami operator}

Inserting the heat kernel (\ref{heatkernel21}) into Eq.~(\ref{radialFractionalLaplaceBeltrami}), we can now compute the inner integral 
\begin{equation}
J_\tau(s,s')\;=\;\int_{|s-s'|}^{s+s'} \d s'' \,\sinh\Bigl( \frac{s''}{L} \Bigr)\,\, K_\tau(s'')\;=\; \frac{e^{-\frac{(s-s')^2}{4 \tau
   }}-e^{-\frac{(s+s')^2}{4 \tau }}}{4\,L\, \pi ^{3/2}\,
   \sqrt{\tau }}\,e^{-\frac{\tau}{L^2}}\,.
\end{equation}
Carrying out the outer integration 
\begin{equation}
J(s,s')\;=\;\int_0^\infty \, \d \tau\, \tau^{-1-\alpha}\, J_\tau(s,s') 
\end{equation}
in Eq.~(\ref{heatkernel21}) we find that
\begin{equation}
\fl
J(s,s')\;=\;
\frac{2^{\alpha-\frac12}}{L^{\alpha+\frac32}\pi^{\frac{3}{2}}}\,
\Bigl( \bigl  |s-s'\bigr|^{-\frac12-\alpha}K_{\frac12+\alpha}\bigl(\frac{|s-s'|}{L}\bigr)
-\bigl(s+s'\bigr)^{-\frac12-\alpha}K_{\frac12+\alpha}\bigl(\frac{s+s'}{L}\bigr) \Bigr)\,,
\end{equation}
where $K_\nu(z)$ denotes the modified Bessel function. Hence we arrive at the kernel representation of the fractional Laplace-Beltrami operator
\begin{equation}
\label{mainResult1}
\fl
\eqalign{
\bigl[\quabla_g^\alpha f \bigr](s)  
\;=\; & -\frac{2^{\alpha+\frac12}}{L^{\alpha+\frac12}\sqrt{\pi}\,\Gamma(-\alpha)}
 \int_0^\infty \d s'\, \frac{\sinh{\bigl(\frac{s'}{L}}\bigr)}{\sinh{\bigl(\frac{s}{L}}\bigr)}\,
 \Bigl( \bigl  |s-s'\bigr|^{-\frac12-\alpha}K_{\frac12+\alpha}\bigl(\frac{|s-s'|}{L}\bigr)
\\ & 
\hspace{40mm}
-\bigl(s+s'\bigr)^{-\frac12-\alpha}K_{\frac12+\alpha}\bigl(\frac{s+s'}{L}\bigr) \Bigr)
 \Bigl(f(s') \,-f(s) \Bigr)
}
\end{equation}
which is the first main result of this paper.
\subsection{Special limits of the kernel representation}
To cross-check our result in~(\ref{mainResult1}) we carried out the following limits:
\begin{itemize}
\item 
In the flat-space limit $L\to \infty$ the arguments of the Bessel functions become small and can be approximated by $K_\nu(z) \;\approx\; z^{-\nu}\, 2^{\nu-1} \, \Gamma(\nu)$ so that Eq.~(\ref{mainResult1}) reduces to
\begin{equation}
\fl
\bigl[\quabla^\alpha f \bigr](s)  = -\frac{2^{2\alpha}\,\Gamma\bigl(\alpha+\frac{1}{2}\bigr)}{\sqrt\pi\,\Gamma\bigl(-\alpha\bigr)}
 \int_0^\infty \d s'\, \frac{s'}{s}\,
 \Bigl( \bigl|s-s'\bigr|^{-1-2\alpha}-\bigl(s+s'\bigr)^{-1-2\alpha} \Bigr)
  \Bigl(f(s') \,-f(s) \Bigr).
\end{equation}
\item 
In the short-range limit $\alpha \to 1$ the fractional Laplace-Beltrami operator reduces to the non-fractional one (see~\ref{AppendixShortRange}):
\begin{equation}
\lim_{\alpha\to 1}\bigl[\quabla_g^\alpha f \bigr](s) \;=\;
\bigl[\quabla_g f \bigr](s) \;=\;
f''(s) + \frac 2 L  \coth\bigl(\frac{s}{L}\bigr)f'(s) 
\end{equation}
\item
In the combined limit $L \to \infty$ and $\alpha \to 1$ one consistently recovers the ordinary d'Alembert operator
\begin{equation}
\lim_{L\to\infty,\alpha\to 1}\bigl[\quabla_g^\alpha f \bigr](s)\;=\; 
\bigl[\quabla \bigr](s)\;=\; f''(s) + \frac{2}{s}f'(s)
\end{equation}
\end{itemize}
%
\section{Fractional Green's function}
%
In this section our aim is to compute the two-point propagator $G_m^\alpha(\xvec,\yvec)$ of the fractional KGE on AdS$_{2+1}$. Being a maximally symmetric space, we expect this function to depend only on the geodesic distance $s(\xvec,\yvec)$, obeying the differential equation
\begin{equation}
(\quabla_g^\alpha-m^2) G_m^\alpha(\xvec,0) = \delta^3(\xvec)\,.
\end{equation}
In terms of geodesic distances $s =s(\xvec)$, we can write equivalently
\begin{equation}
\label{evp}
(\quabla_g^\alpha-m^2) G_m^\alpha(s) = \frac{\delta(s)}{2\pi L^2 \sinh^2(\frac{s}{L})}\,.
\end{equation}
%
\subsection{Inverse eigenmode decomposition}
In the following we determine the function $G_m^\alpha(s)$ without using the explicit kernel representation derived in the previous section. To this end we first solve the eigenvalue problem
\begin{equation}
\quabla_g^\alpha\,\psi_\mu(s) \;=\; -\mu^\alpha  \, \psi_\mu(s)\,,
\end{equation}
finding that the real-valued eigenfunctions of the fractional Laplace-Beltrami operator are given by
\begin{equation}
\label{eigenfunctions}
\psi_\mu (s) \;=\;  \frac{\sin (s \sqrt{\mu-L^{-2}})}{\sinh(\frac s L)}\,.
\end{equation}
These eigenmodes are oscillatory above the Breitenloher-Freedman bound $\mu > \frac{1}{L^2}$. Note that the result is consistent with the composition law $\quabla^\alpha \quabla^\beta\;=\;-\quabla^{\alpha+\beta}$ in Eq.~(\ref{CompositionLaw}).

The next important step is to realize that the right hand side of of Eq.~(\ref{evp}) can be expanded in terms of these eigenfunctions by
\begin{equation}
\frac{\delta(s)}{2 \pi L^2 \,\sinh^2\bigl(\frac s L \bigr)} \;=\; \frac{1}{4\pi^2L}\int_{L^{-2}}^\infty \d \mu \, \psi_\mu (s)\,.
\end{equation}
Therefore, Eq.~(\ref{evp}) can be recast as
\begin{equation}
\label{evp2}
(\quabla_g^\alpha-m^2) G_m^\alpha(s)\;=\;\frac{1}{4\pi^2L}\int_{L^{-2}}^\infty \d \mu \, \psi_\mu (s)\,.
\end{equation}
For this reason we arrive at the formal solution
\begin{equation}
\label{G}
\eqalign{
G_m^\alpha(s)\;&=\;(\quabla_g^\alpha-m^2)^{-1} \,\frac{1}{4\pi^2L}\int_{L^{-2}}^\infty \d \mu \, \psi_\mu (s) \\
&=\;\frac{1}{4\pi^2L}\int_{L^{-2}}^\infty \d \mu \, (\quabla_g^\alpha-m^2)^{-1} \,\psi_\mu (s) \\
&=\;-\frac{1}{4\pi^2L}\int_{L^{-2}}^\infty \d \mu \,\frac{1}{\mu^{\alpha}+m^2} \psi_\mu (s)
\,.}
\end{equation}
%
\subsection{Massless case $m=0$}
%
\begin{figure}
\centering\includegraphics[width=100mm]{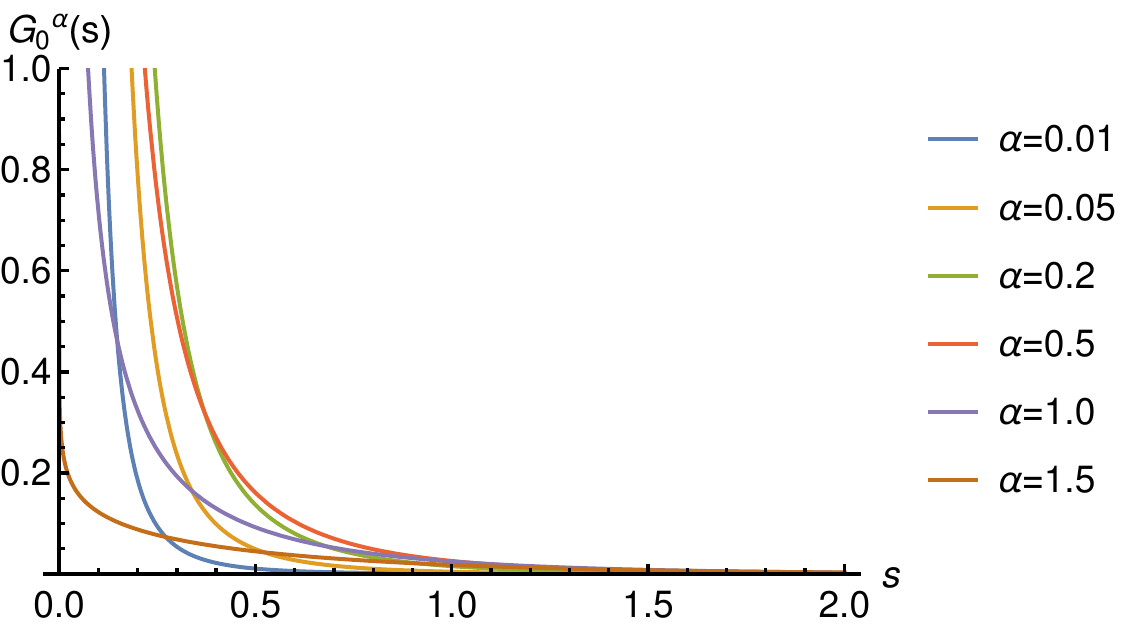}
\caption{Massless Green's function $G_0^\alpha(s)$ for various values of $\alpha$.\label{fig1}}
\end{figure}

Setting $m=0$ the integral~(\ref{G}) can be evaluated explicitly, leading to our second main result, namely, the massless propagator of the fractional wave equation in AdS$_{2+1}$:
\begin{equation}
\label{mainResult2}
G_0^\alpha(s) \;=\; - \frac{\bigl(L s \bigr)^{\alpha-\frac 12}}{L^2 \,\pi^{\frac 3 2\,}2^{\alpha+\frac12}\,\Gamma(\alpha)}
\,\cdot\, \frac{K_{\alpha-\frac32}\bigl(\frac s L \bigr) }{\sinh\bigl(\frac s L \bigr)}
\end{equation}
%
%
which is plotted in Fig.~\ref{fig1}. This function has the following properties:
\begin{itemize}
\item 
The short-range limit reads
\begin{equation}
\lim_{\alpha\to 1}G_0^\alpha(s)\;=\;-\frac{e^{-\frac s L}}{4\pi L \, \sinh(\frac s L )}\,.
\end{equation}
\item 
The flat-space limit 
\begin{equation}
\lim_{L \to \infty}G_0^\alpha(s) \;=\; -\frac{s^{2\alpha-3}\,\Gamma(\frac32 - \alpha)}{2^{2\alpha}\,\pi^{\frac 32}\,\Gamma(\alpha)}\;=\;-\frac{\Gamma(2-2\alpha)\,\sin(\pi\alpha)}{2 \,\pi^2 \, s^{3-2\alpha}}
\end{equation}
is valid only for $\alpha<\frac32$. 
\item
In the large-distance limit $s \to \infty$ we find the approximation
\begin{equation}
G_0^\alpha(s) \;\stackrel{s\to \infty}{\approx}\; -\frac{e^{-s/L} \,L^{\alpha-2}\, s^{\alpha-1}}{2^{1+\alpha}\,\pi\,\Gamma(\alpha) \, \sinh\bigl(\frac s L \bigr)}
\end{equation}
\item 
In the small-distance limit $s \to 0$ the Green's function behaves as
\begin{equation}
G_0^\alpha(s) \;\stackrel{s\to 0}{\approx}\; \left\{
\begin{array}{ll}
-\frac{s^{2\alpha-3}\, \Gamma(\frac32-\alpha)}
      {2^{2\alpha} \, \pi^{3/2}\,\Gamma(\alpha)}\,
& \quad \text{if }\alpha<\frac 32 \\[2mm]
-\frac{L^{2\alpha-3}\, \Gamma(\alpha-\frac32)}
      {8\,\pi^{3/2}\,\Gamma(\alpha)} & \quad \text{if }\alpha>\frac 32
\end{array}
\right.
\end{equation}
\item
We accidentally discovered an exact identity between $G_0^\alpha$ and $G_0^{3-\alpha}$ which is valid in the range $\frac 12 < \alpha < \frac32$:
\begin{equation}
G_0^{3-\alpha}(s) \;=\; \frac{2^{2\alpha-3}\, L^{3-2\alpha}\, s^{3-2\alpha}\, \Gamma(\alpha)}{\Gamma(3-\alpha)}\, G_0^\alpha(s)
\end{equation}
\end{itemize}
%

\subsection{Massive case}
%
For $m>0$ the propagator (\ref{mainResult2}) cannot be evaluated in a closed form. The only exception is the short-range limit $\alpha \to 1$ in which we find the expression
\begin{equation}
G_m(s) \;=\; -  \frac{\exp\bigl(-s\sqrt{\frac 1 {L^2}+m^2}\bigr)}{4 \pi L\,\sinh(\frac s L)}\,.
\end{equation}
This results coincides with the known propagator of the KGE on AdS$_{2+1}$, see e.g.~\cite{ammon2015gauge}. 

For $\alpha<1$ we can expand the integrand as a power series in the mass parameter
\begin{equation}
 \frac{1}{\mu^{\alpha}+m^2} \;=\; \sum_{n=1}^\infty (-m^2)^{n-1}\,  \mu^{-\alpha n} 
\end{equation}
allowing us to formally express the massive Green's function as a series
\begin{equation}
G_m^\alpha(s) \;=\; \sum_{n=1}^\infty (-m^2)^{n-1} G_0^{\alpha n}(s)
\end{equation}
leading to our third main result:
\begin{equation}
\label{MainResult3}
G_m^\alpha(s)\;=\;
-\sum_{n=1}^\infty \frac{(-m^2)^{n-1}\bigl(L s \bigr)^{\alpha n-\frac 12}K_{\alpha n-\frac32}\bigl(\frac s L \bigr)}{L^2 \,\pi^{\frac 3 2\,}2^{\alpha n+\frac12}\,\Gamma(\alpha n)\sinh\bigl(\frac s L \bigr)}\,.
\end{equation}
As shown in~\ref{AppendixConvergence}, this series converges for
\begin{equation}
\label{convergence}
|m^2| < L^{-2\alpha}\,.
\end{equation}
On the one hand, this result suggests that for $m>0$ a convergent flat-space limit $L \to \infty$ does not exist. On the other hand, the result also suggests that a negative squared mass is formally still allowed, provided that the Breitenloher-Freedman bound ($m^2>-L^{-2}$ in the short-range case) is replaced by $m^2>-L^{-2\alpha}$ in the fractional case.

\begin{figure}
\centering\includegraphics[width=100mm]{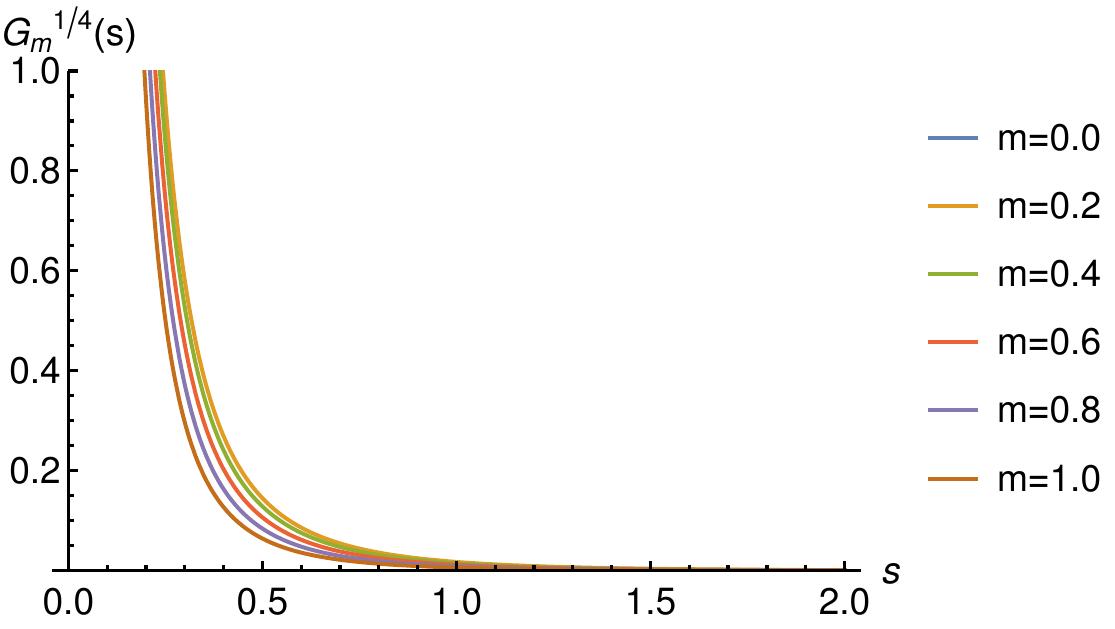}
\caption{\label{fig2}Fractional Green's function $G_m^\alpha(s)$ for $\alpha=\frac14$ and various values of $m$.}
\end{figure}
In Fig.~\ref{fig2} the massive Green's function is shown for a fixed fractional parameter $\alpha=\frac14$, a fixed AdS radius $L=1$ and various masses $m$  ranging from zero up to the convergence limit. As can be seen, the function depends only weakly on the mass parameter.

\section{Numerical verification}

In order to verify the main results of this paper, we performed numerical tests by applying the kernel representation~(\ref{mainResult1}) of the fractional Laplace-Beltrami operator to the eigenfunctions~(\ref{eigenfunctions}) and to the Green's functions (\ref{mainResult2}) and (\ref{MainResult3}). To this end we have to numerically evaluate the integral
\begin{equation}
\label{integral}
\fl
\eqalign{
\bigl[\quabla_g^\alpha f \bigr](s)  
\;=\; & A
 \int_0^\infty \d s'\, \frac{\sinh{\bigl(\frac{s'}{L}}\bigr)}{\sinh{\bigl(\frac{s}{L}}\bigr)}\,
 \Bigl( \bigl  |s-s'\bigr|^{-\frac12-\alpha}K_{\frac12+\alpha}\bigl(\frac{|s-s'|}{L}\bigr)
\\ & 
\hspace{32mm}
-\bigl(s+s'\bigr)^{-\frac12-\alpha}K_{\frac12+\alpha}\bigl(\frac{s+s'}{L}\bigr) \Bigr)
 \Bigl(f(s') \,-f(s) \Bigr)
}
\end{equation}
with the prefactor $A=-\frac{2^{\alpha+\frac12}}{L^{\alpha+\frac12}\sqrt{\pi}\,\Gamma(-\alpha)}$. However, this is technically difficult because the integrand is singular at $s' \approx s$. Moreover, the Bessel functions are sometimes not properly evaluated for very large arguments. Thus it turned out that it is advisable to split the integration range into four parts:
\begin{itemize}
 \item $[0,s-\epsilon]$: Standard numerical integration of~(\ref{integral}).
 \item $[s-\epsilon,s+\epsilon]$: The part close to the singularity is integrated analytically, keeping only the leading terms in $\epsilon$. This contribution is given by
\begin{equation}
 - \epsilon^{2-2\alpha}\frac{\alpha\Gamma(2\alpha)\sin(\pi\alpha)}{L \pi (\alpha-1)}\Bigl( 2 \coth\bigl(\frac{s}{L}\bigr) f'(s) + L f''(s) \Bigr)\,.
\end{equation}
\item $[s+\epsilon,s_c]$:  Standard numerical integration of~(\ref{integral}) up to an upper cutoff at $s_c$.
\item $[s_c,\infty)$: Numerical integration of the integrand, where the Bessel functions are replaced by their approximations for large arguments.
\end{itemize}
\begin{figure}
\centering\includegraphics[width=80mm]{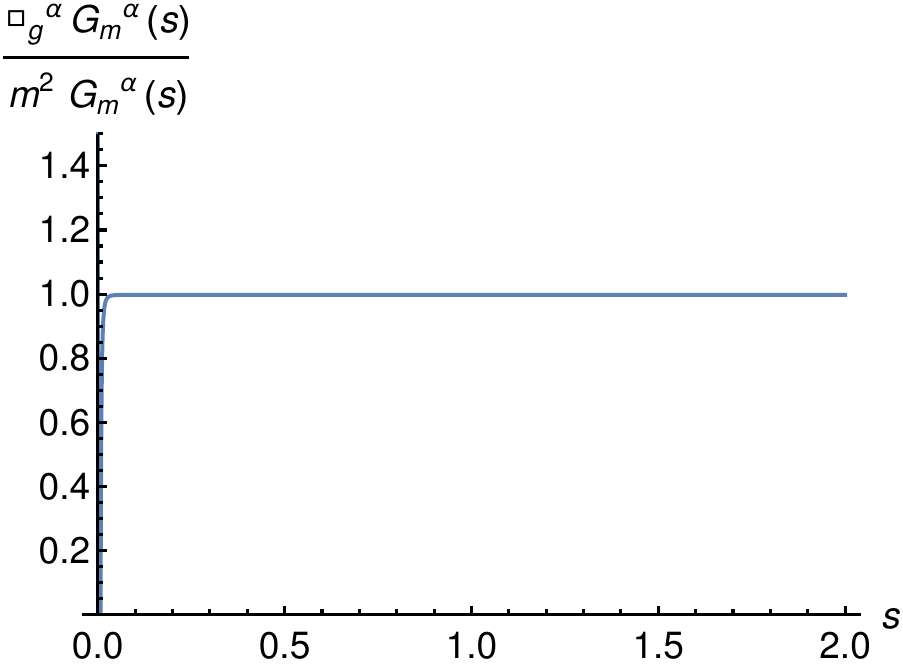}
\caption{\label{fig3}Numerical verification of the massive propagator $G_m^\alpha(s)$ in Eq.~(\ref{MainResult3}) using the kernel representation of $\square_g^\alpha$ in Eq.(\ref{mainResult1}).
} 
\end{figure}

Using the parameters $\epsilon=0.001$ and $s_c=500$ we could confirm our results with an accuracy of about three digits. As an example, let us verify the defining relation
\begin{equation}
\label{evp2}
(\quabla_g^\alpha-m^2) G_m^\alpha(s) = \frac{\delta(s)}{2\pi L^2 \sinh^2(\frac{s}{L})}\,.
\end{equation}
in Eq.~(\ref{evp}) with the series representation of $G_m^\alpha(s)$ in Eq.~(\ref{MainResult3}), using the numerical integration scheme described above. The resulting data is shown in Fig.~\ref{fig3}, where we plotted the numerically computed value of $\square_g^\alpha G^\alpha_m(s)$ divided by $m^2 G^\alpha_m(s)$. For $s>0$, where the r.h.s. of (\ref{evp2}) vanishes, this quotient is expected to be equal to 1. As can be seen, despite the complexity of the formulas, the numerical results confirm this expected behavior accurately. Only for small values of $s$ there are certain deviations, reflecting the presence of the $\delta$-function on the r.h.s. of Eq. (\ref{evp2}) at $s=0$.  

\section{Conclusions}

In this work we have investigated the fractional Klein-Gordon equation and its massless special case, the wave equation, on a 2+1-dimensional AdS space. To this end we proposed a systematic method based on Bochner's approach which allows one to ``fractionalize'' a given equation of motion independent of the background metric.

Our main results are the kernel representation of the Laplace-Beltrami operator~(\ref{mainResult1}) on AdS$_{2+1}$, the massless fractional Green's function~(\ref{mainResult2}) of the wave equation, and the massive Green's function of the fractional Klein-Gordon equation in a series representation~(\ref{MainResult3}). All results are expressed covariantly in terms of geodesic distances. We have restricted the analysis to the simplest non-trivial case of 2+1 dimensions, but in principle it should be possible to obtain similar results in arbitrary dimensions.

Surprisingly our findings suggest that for the fractional Klein-Gordon equation with non-vanishing mass a Green's function does not exist on a flat 2+1-dimensional Minkowski space, but apparently it does exist on AdS$_{2+1}$, provided that the mass $|m^2|<L^{-2\alpha}$ is sufficiently small, cf. Eq.~(\ref{MainResult3}). This leads to the conjecture that massive scalar fields with a fractional dynamics cannot exist in flat spaces but only in spaces with negative curvature. We have therefore identified a phenomenon which relies on the negative curvature of AdS spaces and does not have a zero-curvature limit. Apparently the AdS space offers enough room for a point-like source to spread according to a fractional dynamics while a flat space is too tight. 

With regard to the AdS/CFT correspondence, important questions arise:  Is it still possible to define bulk-to-boundary and boundary-to-boundary correlators in the usual way? Is the induced dynamics at the boundary still conformal? As the long-range interactions decay according to a power law, they do not introduce an additional scale that would rule out a conformal behavior at the boundary. A promising starting point will be to identify the action functional giving rise to the fractional KGE in AdS$_{2+1}$, as well as to analyze the asymptotic behavior of fractional scalar fields near the AdS boundary and to compare it with standard results in AdS/CFT, cf. \cite{ammon2015gauge}. We expect the present paper to be a basis for further studies in this direction.

\appendix

\section{Non-fractional AdS Green's function in general dimensions}
\label{AppendixEquivalence}
%
In the literature~(see e.g.~\cite{ammon2015gauge}) one finds that the scalar propagator in arbitrary dimensions is given in terms of a hypergeometric function by the expression
\begin{equation}
G_\Delta(s) \;=\; \frac{C_\Delta}{2^\Delta (2\Delta-d)}\,\xi^\Delta \cdot {}_2 F_1\Bigl( \frac{\Delta}{2}, \frac{\Delta+1}{2};\, \Delta-\frac{d}{2}+1;\, \xi^2 \Bigr)\,.
\end{equation}
In 2+1 dimensions this result is compatible with (\ref{nonfractionalMassiveGreens}) if we choose
\begin{equation}
\eqalign{
d &\;=\; 2\\
\Delta &\;=\; \frac{d}{2}+\sqrt{\frac{d^2}{4}+m^2L^2}  \\
\xi &\;=\; \frac{1}{\cosh\bigl(\frac{s}{L}\bigr)} \\
C_\Delta &\;=\; \frac{\Gamma(\Delta)}{\pi^{d/2}\Gamma(\Delta-\frac d 2)} 
}
\end{equation}
%
\section{Radial kernel representation}
\label{AppendixKaellenFormula}
%
In this section we show how a kernel integral on AdS$_{2+1}$ can be represented in geodesic distances, generalizing known methods in flat spaces. Starting point is a Cartesian kernel representation
\begin{equation}
g(\xvec) \;=\; \int \d^3 y \, \sqrt{|g|}\, K(\xvec,\yvec) \, f(\yvec)
\end{equation}
in a certain coordinate systems $\vec x=(x^0,x^1,x^2)$ and $\vec y=(y^0,y^1,y^2)$. We assume that both $K$ and $f$ depend only on geodesic distances, i.e.,
\begin{equation}
g\bigl(s(\xvec)\bigr) \;=\; \int \d^3 y \, \sqrt{|g|}\, K\bigl(s(\xvec,\yvec)\bigr) \, f\bigl(s(\yvec)\bigr)\,.
\end{equation}
Inserting a unit factor, namely, $1 = \int \d s' \int \d s'' \, \delta\bigl( s''-s(\xvec,\yvec) \bigr)\, \delta\bigl( s'-s(\yvec) \bigr)$, we obtain
\begin{equation}
\label{gs}
\fl
\eqalign{
g(s) 
&= \int \d^d y \,\sqrt{|g|}\,  \int \d s' \int  \d s'' \, \delta\Bigl( s''-s(\xvec,\yvec) \Bigr)\, \delta\Bigl( s'-s(\yvec) \Bigr)\, K\bigl(\underbrace{s(\xvec,\yvec)}_{=s''}\bigr) \, f \bigl(\underbrace{s(\yvec)}_{=s'}\bigr) \\
&=
\int \d s' \int \d s'' \, \underbrace{\int \d^d y \,\sqrt{|g|}\,  \delta\Bigl( s''-s(\xvec,\yvec) \Bigr)\, \delta\Bigl( s'-s(\yvec) \Bigr)}_{=\chi(s,s',s'')}\, K\bigl(s''\bigr) \, f \bigl(s'\bigr)\,,
}
\end{equation}
where $\xvec$ is a freely point in distance $s$ from the origin. What remains is to determine $\chi(s,s',s'')$ which depends only on the structure of the manifold. To this end one has to rewrite~(\ref{gs}) in a specific coordinate system of AdS$_{2+1}$, for example, in global coordinates $t,r,\phi$. Due to radial symmetry, the integration over $\phi$ drops out, giving a factor $2\pi$. The remaining two integrations can be evaluated by using the rule $\delta(y-g(x))=\frac{1}{|g'(x_0)|}\delta(x-x_0)$, where $x_0$ is the root of $y-g(x)=0$. Elementary calculations (not shown here) lead to the result
\begin{equation}
\label{chiresult}
\chi(s,s',s'') \;=\; \left\{ \begin{array}{ll} 2\pi L \frac{\sinh(\frac{s'}{L})\sinh(\frac{s''}{L})}{\sinh(\frac{s}{L})} & \text{if } |s-s'|<s''<|s+s'|\\ 0 & \text{otherwise.} \end{array} \right.
\end{equation}
Alternatively, this expression can be derived by a coordinate transformation $(s',s'') \leftrightarrow (t,r)$. In this context it is interesting to note that the Jacobian can be expressed in terms of the well-known Källén function~\cite{kallen1964elementary} generalized to AdS:
\begin{equation}
\eqalign{
\Lambda(s,s',s'')\;=\;16&\, L^4\, \sinh\Bigl(\frac{s+s'+s''}{2L}\Bigr)\sinh\Bigl(\frac{s+s'-s''}{2L}\Bigr)\\ &\times\;
\sinh\Bigl(\frac{s-s'+s''}{2L}\Bigr)\sinh\Bigl(\frac{s-s'-s''}{2L}\Bigr)
}
\end{equation}
To our knowledge this generalization has not been known so far. This function has to be positive, explaining immediately the bounds in (\ref{chiresult}).

\section{Short-range limit in the kernel representation}
\label{AppendixShortRange}

In this Appendix we show that the kernel representation~(\ref{mainResult1}) of the fractional Laplace-Beltrami operator, given by
\begin{equation}
\label{toprove}
\fl
\eqalign{
\bigl[\quabla_g^\alpha f \bigr](s)  
\;=\; & -\frac{2^{\alpha+\frac12}}{L^{\alpha+\frac12}\sqrt{\pi}\,\Gamma(-\alpha)}
 \int_0^\infty \d s'\, \frac{\sinh{\bigl(\frac{s'}{L}}\bigr)}{\sinh{\bigl(\frac{s}{L}}\bigr)}\,
 \Bigl( \bigl  |s-s'\bigr|^{-\frac12-\alpha}K_{\frac12+\alpha}\bigl(\frac{|s-s'|}{L}\bigr)
\\ & 
\hspace{40mm}
-\bigl(s+s'\bigr)^{-\frac12-\alpha}K_{\frac12+\alpha}\bigl(\frac{s+s'}{L}\bigr) \Bigr)
 \Bigl(f(s') \,-f(s) \Bigr)
}
\end{equation}
reduces neatly to the short-range differential equation in the limit $\alpha\to 1$. Starting point is the observation that $\Gamma(-\alpha) \approx \frac{1}{\alpha-1}$ diverges in this limit, eliminating all non-divergent contributions of the integral. Since the divergent contributions are peaked around $s'\approx s$, we can limit the integration range to $s'\in[s-\epsilon,s+\epsilon]$ and finally show that in the limit $\alpha\to 1$ the result becomes independent of $\epsilon$.

Since the singularity at $s'\approx s$ is produced only by the first Bessel function in~(\ref{toprove}), we can discard the second Bessel function which depends on $s+s'$. The remaining factors can be Taylor-expanded in $s'$ around $s$ as
\begin{equation}
\eqalign{
&\frac{\sinh{\bigl(\frac{s'}{L}}\bigr)}{\sinh{\bigl(\frac{s}{L}}\bigr)}
\;=\;
1 + \frac{1}{L}\coth\bigl(\frac{s}{L}\bigr)(s'-s)+\ldots\nonumber \\
&\bigl  |s-s'\bigr|^{-\frac12-\alpha}
K_{\frac12+\alpha}\bigl(\frac{|s-s'|}{L}\bigr)
\;=\; 
\frac{2^{\alpha-\frac12} \,L^{\alpha+\frac12}\,\Gamma(\alpha+\frac12 )}{|s-s'|^{2\alpha+1}} +\ldots
\nonumber \\
& f(s')-f(s) \;=\; (s'-s) f'(s) + \frac12 (s'-s)^2 f''(s)+\ldots
}
\end{equation}
In the integrand only symmetric contributions in $s-s'$ will survive:
\begin{equation*}
\fl
\eqalign{
\lim_{\alpha\to 1}
\bigl[\quabla_g^\alpha f \bigr](s)  
&\;=\; -\lim_{\alpha\to 1}\frac{2^{2\alpha}\Gamma(\alpha+\frac12)}{\sqrt{\pi}\Gamma(-\alpha)}
 \int_{s-\epsilon}^{s+\epsilon} \d s'\, 
 \frac{1}{|s-s'|^{2\alpha+1}}
 \Bigl( 1 + \frac{\coth\bigl(\frac{s}{L}\bigr)(s'-s) }{L}\Bigr)\\
 &\hspace{55mm} \times \;
 \Bigl( (s'-s) f'(s) + \frac{(s'-s)^2}2 f''(s) \Bigr)
 \\
&\;=\; -2
\Bigl( \frac12 f''(s) + \frac 1 L  \coth\bigl(\frac{s}{L}\bigr)f'(s)\Bigr)
\lim_{\alpha\to 1}\,\underbrace{(\alpha-1)
 \int_{s-\epsilon}^{s+\epsilon} \d s'\, 
 \frac{1}{|s-s'|^{2\alpha-1}}}_{=\epsilon^{2-2\alpha}} \\
& \;=\; f''(s) + \frac 2 L  \coth\bigl(\frac{s}{L}\bigr)f'(s) \;=\;
\bigl[\quabla_g f \bigr](s)  \,.
}
\end{equation*}
%

\section{Convergence radius in the massive case}
\label{AppendixConvergence}

In the following we sketch a possible way to determine the convergence radius of the massive Green's function~(\ref{MainResult3})
\begin{equation}
G_m^\alpha(s) \;=\;
-\frac{1}{L^2\pi^{\frac 3 2}\sinh\bigl(\frac s L \bigr)}
\sum_{n=1}^\infty (-m^2)^{n-1}\frac{\bigl(L s \bigr)^{\alpha n-\frac 12}K_{\alpha n-\frac32}\bigl(\frac s L \bigr)}{2^{\alpha n+\frac12}\,\Gamma(\alpha n)}\,.
\end{equation}
Let us first consider the case that $s>0$ is real. Since the modified Bessel function is positive in this case, we can estimate
\begin{equation}
|G_m^\alpha(s)| \leq
\frac{1}{L^2\sinh\bigl(\frac s L \bigr)}
\sum_{n=1}^\infty |m^2|^{n-1} \frac{\bigl(L s \bigr)^{\alpha n-\frac 12}K_{\alpha n-\frac32}\bigl(\frac s L \bigr)}{\pi^{\frac 3 2\,}2^{\alpha n+\frac12}\,\Gamma(\alpha n)\sinh\bigl(\frac s L \bigr)}
\end{equation}
For $\nu>0$ and $z\geq 0$ the modified Bessel function obeys the inequality
\begin{equation}
K_\nu(z) \,\leq\, \frac12 \frac{\Gamma(\nu)}{(z/2)^\nu}\,.
\end{equation}
Thus, for $s>0$ the convergence of
\begin{equation}
\sum_{n=N}^\infty\frac{m^{2n}(Ls)^{\alpha n}\Gamma(\alpha n-\frac32)}{2^{\alpha n}\Gamma(\alpha n)(\frac s {2L})^{\alpha n}}
\;=\;
\sum_{n=N}^\infty |m^2|^{n} L^{2\alpha n} \frac{\Gamma(\alpha n-\frac 32)}{\Gamma(\alpha n)}
\end{equation}
implies the convergence of $G_m^\alpha(s)$. Here we omitted the first $N-1$ terms, where $N$ is the smallest integer such that the order of the Bessel function $\alpha N-\frac 3 2$ is positive. Since $\frac{\Gamma(\alpha n-\frac 32)}{\Gamma(\alpha n)}$ converges quickly to $(\alpha n)^{-3/2}$, one obtains essentially an expression involving a polylogarithm $\text{Li}_{3/2}\bigl(|m^2|L^{2\alpha}\bigr)$, the power series of which is convergent only for arguments smaller than 1. Thus the convergence radius is equal or larger than $|m^2| < L^{-2\alpha}$. A numerical survey suggests that this is in fact the actual convergence radius of the series expansion~(\ref{MainResult3}) and that this holds even for imaginary values of $s$.

\section*{References}
\bibliographystyle{unsrt}
\bibliography{paper}
\end{document}